\documentclass[twocolumn]{bmcart}% uncomment this for twocolumn layout and comment line below
\usepackage{amsthm,amsmath}
\usepackage[utf8]{inputenc} %unicode support

\usepackage{graphicx}
\newtheorem{lemma}{Lemma}%[section]
\newtheorem{proposition}{Proposition}%[section]
\newtheorem{theorem}{Theorem}%[section]

\startlocaldefs
\endlocaldefs

\begin{document}

%%% Start of article front matter
\begin{frontmatter}

\begin{fmbox}
\dochead{Research}


\title{Generating Normal Networks via Leaf Insertion and Nearest Neighbor Interchange}

%%%%%%%%%%%%%%%%%%%%%%%%%%%%%%%%%%%%%%%%%%%%%%
%%                                          %%
%% Enter the authors here                   %%
%%                                          %%
%% Specify information, if available,       %%
%% in the form:                             %%
%%   <key>={<id1>,<id2>}                    %%
%%   <key>=                                 %%
%% Comment or delete the keys which are     %%
%% not used. Repeat \author command as much %%
%% as required.                             %%
%%                                          %%
%%%%%%%%%%%%%%%%%%%%%%%%%%%%%%%%%%%%%%%%%%%%%%

\author[
   addressref={aff1},                   % id's of addresses, e.g. {aff1,aff2}
   corref={aff1},                       % id of corresponding address, if any
   %noteref={n1},                        % id's of article notes, if any
   email={matzlx@nus.edu.sg}   % email address
]{\inits{LX}\fnm{Louxin} \snm{Zhang}}
%\author[
%   addressref={aff1,aff2},
%   email={john.RS.Smith@cambridge.co.uk}
%]{\inits{JRS}\fnm{John RS} \snm{Smith}}

%%%%%%%%%%%%%%%%%%%%%%%%%%%%%%%%%%%%%%%%%%%%%%
%%                                          %%
%% Enter the authors' addresses here        %%
%%                                          %%
%% Repeat \address commands as much as      %%
%% required.                                %%
%%                                          %%
%%%%%%%%%%%%%%%%%%%%%%%%%%%%%%%%%%%%%%%%%%%%%%

\address[id=aff1]{%                           % unique id
  \orgname{Department of Mathematics, National University of Singapore}, % university, etc
  \street{10 Lower Kent Ridge Road},                     %
  \postcode{119076}                                % post or zip code
  \city{Singapore},                              % city
  \cny{Singapore}                                    % country
}

%%%%%%%%%%%%%%%%%%%%%%%%%%%%%%%%%%%%%%%%%%%%%%
%%                                          %%
%% Enter short notes here                   %%
%%                                          %%
%% Short notes will be after addresses      %%
%% on first page.                           %%
%%                                          %%
%%%%%%%%%%%%%%%%%%%%%%%%%%%%%%%%%%%%%%%%%%%%%%

%\note{Sample of title note}     % note to the article
%\note[id=n1]{Equal contributor} % note, connected to author

\end{fmbox}% comment this for two column layout

%%%%%%%%%%%%%%%%%%%%%%%%%%%%%%%%%%%%%%%%%%%%%%
%%                                          %%
%% The Abstract begins here                 %%
%%                                          %%
%% Please refer to the Instructions for     %%
%% authors on http://www.biomedcentral.com  %%
%% and include the section headings         %%
%% accordingly for your article type.       %%
%%                                          %%
%%%%%%%%%%%%%%%%%%%%%%%%%%%%%%%%%%%%%%%%%%%%%%

\begin{abstractbox}

\begin{abstract} % abstract
\parttitle{Background} %if any
Galled trees are studied as a recombination model in theoretical population genetics.  This class of phylogenetic networks has been generalized to  tree-child networks and other network classes by relaxing a structural condition imposed on galled trees. Although these networks are simple, their topological structures have yet to be fully understood.

\parttitle{Results} %if any
 It is well-known that all phylogenetic trees on $n$ taxa can be generated by the insertion of the $n$-th taxa to each  edge of all the phylogenetic trees on $n-1$ taxa. We prove that all tree-child (resp. normal) networks with $k$ reticulate nodes on $n$ taxa can be uniquely generated via three operations from all the tree-child (resp. normal) networks with $k-1$ or $k$ reticulate nodes on $n-1$ taxa. Applying this result to counting rooted phylogenetic networks, we show that  there are exactly 
$\frac{(2n)!}{2^n (n-1)!}-2^{n-1} n!$ binary phylogenetic networks with one reticulate node on $n$ taxa.

%\parttitle{Conclusions}

\end{abstract}

%%%%%%%%%%%%%%%%%%%%%%%%%%%%%%%%%%%%%%%%%%%%%%
%%                                          %%
%% The keywords begin here                  %%
%%                                          %%
%% Put each keyword in separate \kwd{}.     %%
%%                                          %%
%%%%%%%%%%%%%%%%%%%%%%%%%%%%%%%%%%%%%%%%%%%%%%


% MSC classifications codes, if any
%\begin{keyword}[class=AMS]
%\kwd[Primary ]{}
%\kwd{}
%\kwd[; secondary ]{}
%\end{keyword}

\end{abstractbox}
%
%\end{fmbox}% uncomment this for twcolumn layout

\end{frontmatter}

%%%%%%%%%%%%%%%%%%%%%%%%%%%%%%%%%%%%%%%%%%%%%%
%%                                          %%
%% The Main Body begins here                %%
%%                                          %%
%% Please refer to the instructions for     %%
%% authors on:                              %%
%% http://www.biomedcentral.com/info/authors%%
%% and include the section headings         %%
%% accordingly for your article type.       %%
%%                                          %%
%% See the Results and Discussion section   %%
%% for details on how to create sub-sections%%
%%                                          %%
%% use \cite{...} to cite references        %%
%%  \cite{koon} and                         %%
%%  \cite{oreg,khar,zvai,xjon,schn,pond}    %%
%%  \nocite{smith,marg,hunn,advi,koha,mouse}%%
%%                                          %%
%%%%%%%%%%%%%%%%%%%%%%%%%%%%%%%%%%%%%%%%%%%%%%

%%%%%%%%%%%%%%%%%%%%%%%%% start of article main body
% <put your article body there>

%%%%%%%%%%%%%%%%
%% Background %%
%%
\section*{Background}
Phylogenetic networks have been used to date both vertical and horizontal genetic transfers  in evolutionary genomics and population genetics in the past two decades \cite{Doolittle_99, Gusfield_book, Lake_99}. A rooted phylogenetic network (RPN) is a directed acyclic digraph in which  all the sink nodes are of indegree 1 and a unique source node is designated as the root, where the former represent a set of taxa (e.g, species, genes, or individuals in a population) and the latter represents the least common ancestor of the taxa.  Moreover, the other nodes in  a RPN are divided into tree nodes and reticulate nodes, where reticulate nodes represent reticulate evolutionary events such as horizontal genetic transfers and genetic recombination. 

The topological properties of RPNs are much more complicated than phylogenetic trees \cite{Gusfield_book,Huson_book,Steel_book}. Therefore,  different mathematical issues arise in the study of RPNs.  First,   phylogenetic reconstruction problems are often NP-hard even for 
trees \cite{Chor_2006,Foulds_1982}. As such, 
a phylogenetic reconstruction method often uses  nearest neighbor interchanges (NNIs) or other  rearrangement operations to search for an optimal  tree or network \cite{Felsenstein,Yu_2014}. 
Recently, different variants of  NNI have been proposed for RPNs \cite{Bordewich_17,Francis_18,Gambette_17,Huber_Linz_16,Huber_16,Janssen_18,Klawitter_18}.

Second, to develop efficient algorithms for NP-complete problems on RPNs,  simple classes of RPNs have been introduced,  including galled trees \cite{Gusfield_04, Wang_01}, 
tree-child networks (TCNs) \cite{Cardona_09b}, normal networks \cite{Willson_07},  reticulation-visible networks \cite{Huson_book} and tree-based networks \cite{Francis_15,Zhang_18} (see also \cite{Steel_book, Zhang_19}).  For instance, a RPN is a TCN if every non-leaf node has a child that is a tree node or a leaf. Although these network classes have been intensively investigated, their topological structures remain unclear 
\cite{Gunawan_19,Steel_book}. How to efficiently enumerate and count normal networks  remains unclear \cite{Bicker_12,Bouvel_18,Cardona_19,Fuchs_18,Semple_15,Steel_06}.

This work makes a contribution to understanding TCNs and normal networks.  It is a well-known fact that  all phylogenetic trees on $n$ taxa can be generated by inserting the $n$-th taxa in every  edge of  all the phylogenetic trees on $n-1$ taxa.
We prove that all TCNs with $k$ reticulate nodes on $n$ taxa can be uniquely generated via three operations from TCNs with $k-1$ or $k$ reticulate nodes on $n-1$ taxa (Theorem~\ref{enumeration_thm}, Section~\ref{sec3}). Using this fact, we obtain recurrence formulas 
for counting TCNs and normal networks (Section~\ref{sec4}). In particular,   simple  formulas are given for the number  of RPNs and normal networks with one reticulate node, respectively.

\section*{Basic notation}
\label{sec2}

\subsection*{RPNs}

\begin{figure}[t!]
            \centering
            \includegraphics[scale = 0.9]{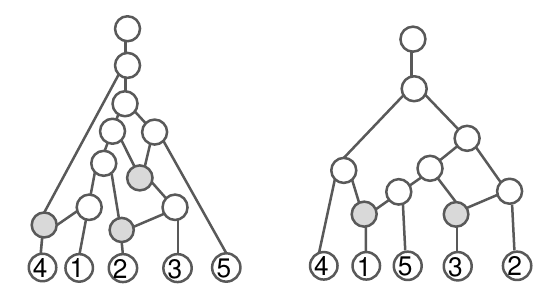}
           \caption{Two tree-child networks on $\{1, 2, 3, 4, 5\}$, where reticulate and tree nodes are drawn as filled and unfilled circles, respectively. Only the right network is normal. Here,  edge downward orientation is omitted.}
            \label{Fig_1}
\end{figure}

 A RPN over a finite set of taxa $X$ is an  acyclic digraph such that:
\begin{itemize}
   \item there is a unique node of indegree 0 and outdegree 1, called the {\it root}; 
   \item there are exactly $|X|$ nodes of outdegree 0 and indegree 1, called the {\it leaves} of the RPN, each being labeled with a unique taxon in $X$;
 \item  each non-leaf/non-root node is either a {\it reticulate node} that is of indegree 2 and outdegree 1,  or a {\it tree node} that is of indegree 1 and outdegree 2; and
\item there are no parallel edges between a pair of nodes.
\end{itemize}
Two RPNs are drawn in Figure~\ref{Fig_1}, where each edge is directed away from the root and both the root and edge orientation are omitted.  For a RPN $N$, we  use ${\cal V}(N)$, ${\cal R}(N)$, 
${\cal T}(N)$ and  ${\cal E}(N)$ to denote the set of all nodes, the set of reticulate nodes and 
the set of tree nodes and the set of directed edges for $N$, respectively.

Let $u\in {\cal V}(N)$ and $v\in {\cal V}(N)$.  The node  $u$ is said to be a {\it parent} (resp. a {\it child}) of  $v$ if $(u, v)\in {\cal E}(N)$ (resp. $(v, u)\in {\cal E}(N)$).  Every reticulate node $r$ has a unique child, named $c(r)$, whereas every tree node $t$ has a unique parent, named $p(t)$.
Furthermore, $u$ is an {\it ancestor} of $v$ or,  equivalently, $v$ is {\it below} $u$ if there is a direct path from  the network root to $v$ that contains $u$.  We say that $u$ and $v$ are {\it incomparable} 
if neither of them is an ancestor of the other. 

Let $e=(u, v)\in {\cal E}(N)$. It is a {\it reticulate edge} if $v$ is a reticulate node and a {\it tree edge} otherwise.
Hence, a tree edge leads to either a tree node or a leaf. 

A {\it phylogenetic tree}  is simply a  RPN with no reticulate nodes.

A {\it TCN} is a  RPN in which every non-leaf node has a child that is a tree node or a leaf or, equivalently, there is a path from every non-leaf node to some leaf that consists only of  tree edges. Both RPNs in Figure~\ref{Fig_1} are tree-child. 

A {\it normal} network is a TCN in which  every reticulate node satisfies the following condition:
\begin{quote}
   ({\bf The normal condition})  The two parents  are incomparable. 
\end{quote}  
The first PRN in Figure~\ref{Fig_1} is not normal, as  a parent of the left most reticulate node is an ancestor of the other in the network.

\section*{Generating TCNs and Normal Networks}
\label{sec3}

\begin{figure}[b!]
            \centering
            \includegraphics[scale = 0.8]{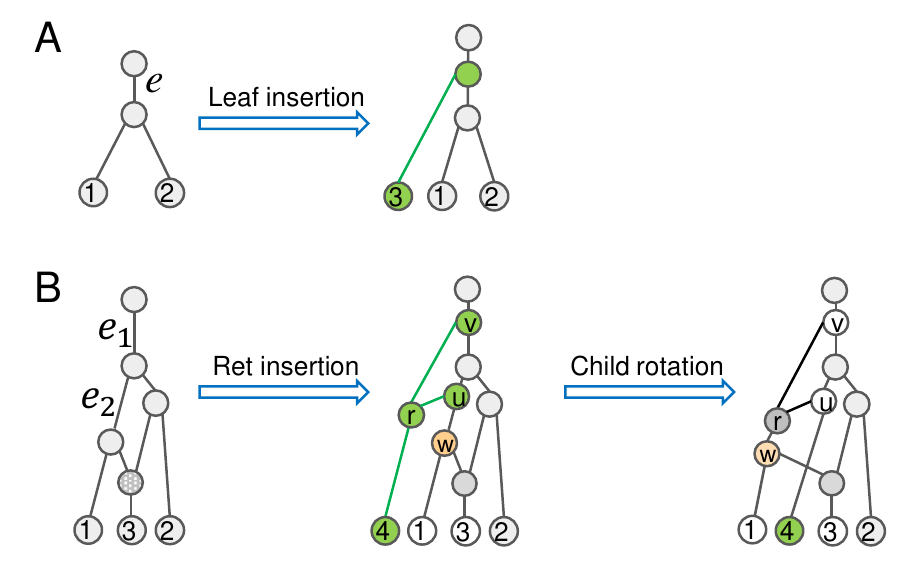}
           \caption{{\bf Insertion and child rotations for tree-child networks}. 
A.   Leaf 3 is attached to a tree edge.  B. The reticulation insertion is applied to attach a new reticulate node $r$ onto two tree edges. The child rotation swaps the tree node $w$ (yellow) and Leaf 4.   Here, green nodes and edges are added nodes.}
            \label{operation_ill}
\end{figure}

 We define the following rearrangement operations for TCNs $N$ on $[1, n]$, which are illustrated in Figure~\ref{operation_ill}:
\begin{itemize} 
\item {\bf Leaf insertion} ~For a tree edge  $e=(u, v)\in {\cal E}(N)$, insert a new node $w$ to subdivide $e$ and attach Leaf $n+1$ below $w$ as its child. The resulting network  is denoted by $\mbox{Leaf-Insert}(N, e, n+1)$, in which  $w$ is a tree node. 
\item {\bf Reticulation insertion}  ~For a pair of tree edges $e_1=(u_1, v_1)$ and $e_2=(u_2, v_2)$ of $N$,  which are not necessarily distinct, insert a new node $w_1$ to subdivide $e_1$ and a new node $w_2$ to subdivide $e_2$, attach a new reticulate  node $r$ as the common child of $w_1$ and $w_2$ and make Leaf $(n+1)$ to be the child of $r$. In this case, we say that 
$r$ {\it straddles} $e_1$ and $e_2$. We  use $\mbox{Ret-Insert}(N, e_1, e_2, n+1)$ to denote the resulting network.   We simply write 
 $\mbox{Ret-Insert}(N, e, n+1)$ if $e_1=e_2=e$.
\item {\bf Child rotation} ~Let $r$ be a reticulate node with parents $u \in {\cal T}(N)$ and $v$. If $u$ is not an ancestor of $v$, exchange the unique child of $r$ and  the other child of $u$. The resulting network is denoted by $\mbox{C-Rotate}(N, u, r)$.
\end{itemize} 

Note that a child rotation is a special case of  the rNNI rearrangement introduced by Gambette et al. in \cite{Gambette_17}.   Let ${\cal TCN}_k(n)$ denote the set of TCNs with $k$ reticulations on $[1, n]$. 
\begin{proposition}
\label{prop1}
 Let $M\in {\cal TCN}_{k}(n)$ and let  $e_1$ and $e_2$ be two tree edges of $M$. Then, 
\begin{eqnarray*}
   \mbox{\rm Ret-Insert}\left(M, e_1, e_2, n+1\right)\in {\cal TCN}_{k+1}(n+1),\\
\mbox{\rm Ret-Insert}(M, e_1, n+1)\in {\cal TCN}_{k+1}(n+1).
\end{eqnarray*}
\end{proposition} 
{\bf Proof}. The second statement is a special case of the first. 
Let $e_1=(u, v)$ and $e_2=(x, y)$. Since $e_1$ and $e_2$  are tree edges, both $v$ and $y$ are  tree nodes or leaves. Let $r$ be the added reticulate node.  Then the parents of $r$ have $v$ and $y$ as their child, respectively,  the nodes $u$ and $x$ have the parents of $r$ as their tree node child; Leaf $n+1$ is the tree child $r$. Additionally,  all the other nodes have the same children as in $M$.  Therefore,   
$\mbox{Ret-Insert}\left(M, e_1, e_2, n+1\right)$  is a TCN.

%The second statement can be proved similarly. 
%$\QED$

\begin{proposition}
 Let $M\in {\cal TCN}_{k+1}(n+1)$. Assume that $r\in {\cal R}(M)$ and its parents are $u$ and $v$
such that $u$ is not an ancestor of $v$ in $M$. Then, 
\begin{eqnarray*}
   \mbox{\rm C-Rotate}\left(M, u, r \right)\in {\cal TCN}_{k+1}(n+1).
\end{eqnarray*}
\end{proposition} 
{\bf Proof}.   Let $M'=\mbox{\rm C-Rotate}\left(M, u, r \right)$.
Since $u$ is a parent of $r$ and $M$ is tree-child,  $u$ is a tree node. Let $w$ be  the other child of  $u$ and let $z$ be the unique child of $r$.  Since $M$ is tree-child, $z$ and $w$ are tree nodes (see Figure~\ref{operation_ill}). The tree node $z$ becomes the child of $u$  Therefore,   every node also has a child that is a tree node in $M'$.

By definition,  $w$ becomes a child of $r$ and  $z$ becomes a tree node child of $u$ in $M'$.    
 If $M'$ contains a directed cycle $C$, $C$ must contain $v$ and $w$, implying that $u$ is an ancestor of $v$ in $M$, a contradiction. Therefore, $M'$ is acyclic and $M' \in  {\cal TCN}_{k+1}(n+1)$.  
%$\QED$

\begin{figure}[t!]
            \centering
            \includegraphics[scale = 0.9]{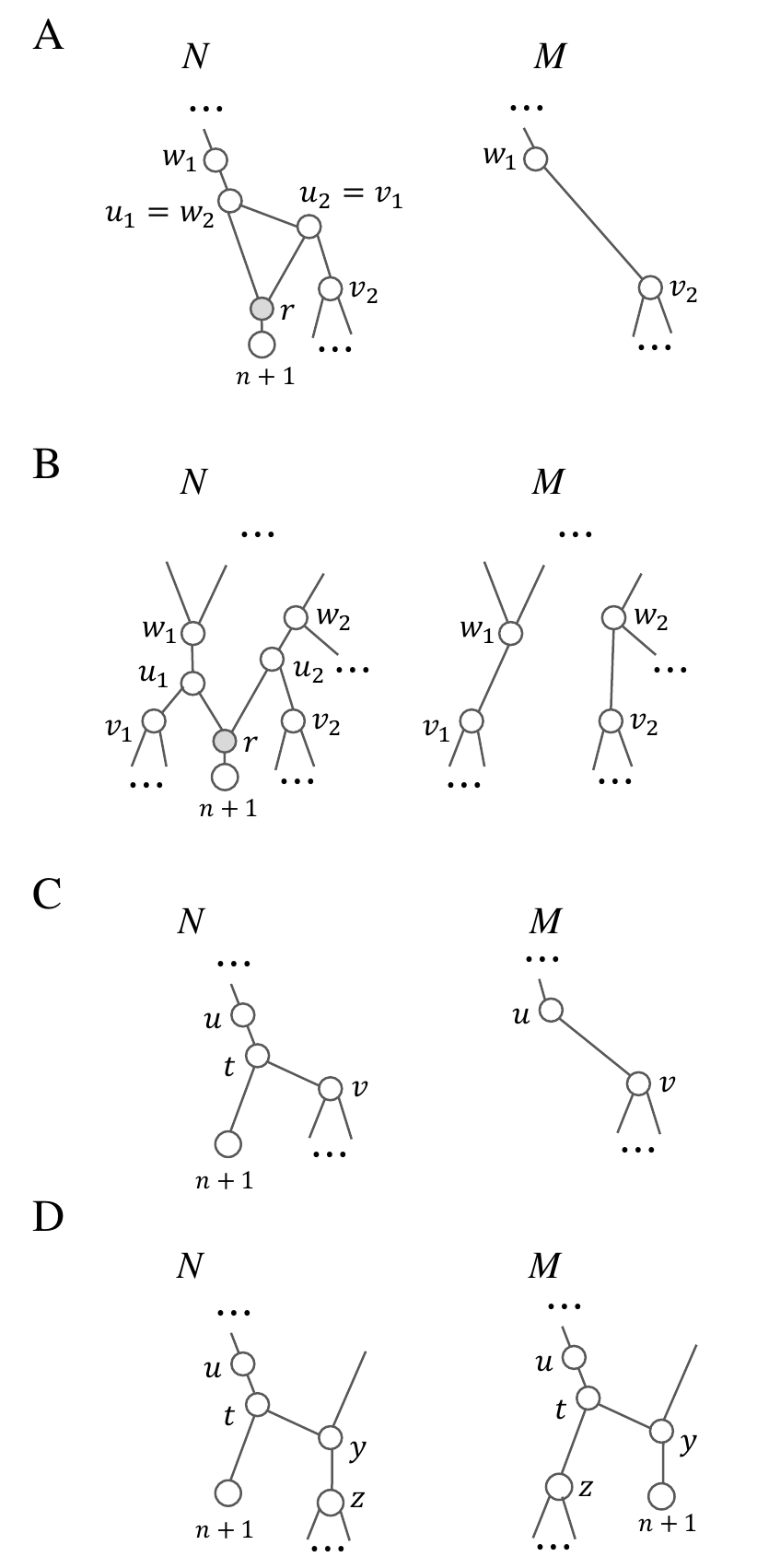}
           \caption{{\bf An illustration of the proof of Proposition~\ref{prop3}}. A.  The reticulate node parent $r$ of 
$n+1$  has  two adjacent  parents. B.  The reticulate node parent $r$ of 
$n+1$  has  two non-adjacent  parents.   C.  The parent and sibling of 
 $n+1$ are both a tree node. D. The parent of  $n+1$ is a tree node, whereas the sibling of $n+1$ is a reticulate node. 
}
            \label{ill_prop3}
\end{figure}

\begin{proposition}
\label{prop3}
 Let $N\in {\cal TCN}_{k+1}(n+1)$.  

{\rm (i)} If Leaf $(n+1)$ is the child of a reticulate node $r$, $N$ can then be obtained from an $M\in {\cal TCN}_{k}(n)$ via  a reticulation insertion. 

{\rm (ii)} If  Leaf $(n+1)$ is the child of a tree node $t$ and the sibling of $n+1$ is also a tree node,   $N$ can then obtained from  an $M\in {\cal TCN}_{k+1}(n)$ via a leaf insertion. 

{\rm (iii)} If   Leaf $(n+1)$ is a child of a tree node $t$ and the sibling of $n+1$  is a reticulate node,
$N$ can then be obtained from an $M\in {\cal TCN}_{k+1}(n+1)$ via a child rotation.  
%That is  $N=\mbox{C-Rotate}\left(M, x, y\right)$,  where $y$ is a reticulate node such that 
% Leaf $(n+1)$ is the child of $y$ and $x$ is a parent of $y$ in $M$. 
\end{proposition}
{\bf Proof.} (i) Let  $r$ have parents  $u_1$ and $u_2$ in $N$. Since $N$ is a TCN, $u_1$ and $u_2$ are tree nodes and so are their children other than $r$.  Let $w_i$  and $v_i$ be the parent  and the child of $u_i$ such that $v_i\neq r$, respectively, for each $i=1, 2$. Since $r$ is a reticulate node, $v_1$ and $v_2$ are tree nodes.  Without loss of generality, we assume that $u_2$ is not the parent of $u_1$. There are two cases for consideration. 

If $u_1$ is the parent of $u_2$, then $u_1=w_2$ and $u_2=v_1$ (Figure~\ref{ill_prop3}A). Removing Leaf $(n+1)$, $u_1$ and $u_2$ (together with incident edges) and adding an edge $e=(w_1, v_2)$ produce a TCN $M$ with  $k$ reticulations such that
 $N=\mbox{Ret-Insert}\left(M, e, n+1\right)$. 

If $u_1$ is not the parent of $u_2$, then, $w_1\neq u_2$ (Figure~\ref{ill_prop3}B). After removing Leaf $n+1$, $u_1$ and $u_2$ (together with incident edges) and adding two edges $e_i=(w_i, v_i)$ ($i=1, 2$), we obtain a TCN $M$ such that $N=\mbox{Ret-Insert}\left(M, e_1, e_2 , n+1\right)$. 

(ii) Let $u$ be the parent of $t$ and let  $v$ be  the sibling of Leaf $n+1$ (Figure~\ref{ill_prop3}C).
By assumption, $v$ is a tree node.  After removing $t$ and Leaf $(n+1)$ (together with incident edges) and adding 
$e=(u, v)$, we obtain a TCN $M\in {\cal TCN}_{k+1}(n)$ such that 
 $N=\mbox{Leaf-Insert}\left(M, (u, v), n+1\right)$. 

(iii) Let $y$ be the sibling of $n+1$ that is a reticulate node (Figure~\ref{ill_prop3}D).  Let $z$ be the child of $y$ and let $M=\mbox{C-Rotate}\left(N, t, y\right)$. Since $z$ is below $y$ and $y$ is below $t$ in $N$, 
neither attaching the tree node $z$ below $t$ nor attaching Leaf $(n+1)$ below $y$ generates  a directed cycle in $M$.  Hence, $M\in {\cal TCN}_{k+1}(n+1)$ in which Leaf $(n+1)$ is the child of a reticulate node $y$ such that $N=\mbox{C-Rotate}\left(M, t, y\right)$. 
%$\QED$

\begin{figure}[b!]
            \centering
\includegraphics[scale = 0.8]{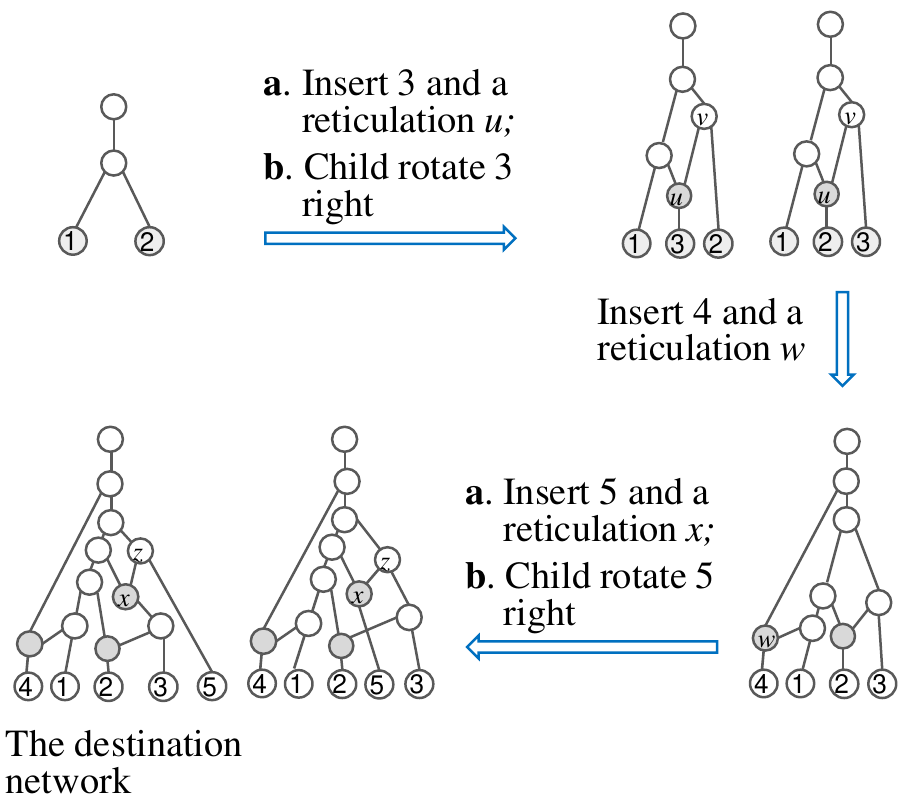}
           \caption{Illustration how to generate the TCN in Figure~1 from a tree on 2 taxa.}
            \label{Example_3}
\end{figure}

\begin{proposition} 
\label{prop4}
Let $N_1, N_2 \in {\cal TCN}_{k}(n)$.
  
 {\rm (i)} $\mbox{\rm Leaf-Insert}\left(N_1, e_1, n+1\right)$ is identical to\\ $\mbox{\rm Leaf-Insert}\left(N_2, e_2, n+1\right)$ iff $N_1=N_2$ and $e_1=e_2$. 

{\rm (ii)} $\mbox{\rm Ret-Insert}\left(N_1, e_1, e_1', n+1\right)$ is identical to\\
 $\mbox{\rm Ret-Insert}\left(N_2, e_2, e'_2, n+1\right)$ iff $N_1=N_2$. 

{\rm (iii)} Assume  the parent of Leaf $n$ is a reticulate node $y_i$ in $N_i$ for $i=1,2$. 
$\mbox{\rm C-Rotate}\left(N_1, x_1, y_1\right)$ is identical to $\mbox{\rm C-Rotate} \left(N_2, x_2, y_2\right) $
 iff $N_1=N_2$. 
\end{proposition}
{\bf Proof.} (i) Let $N_i \in {\cal TCN}_{k}(n)$ and $e_i\in {\cal V}(N_i)$, $i=1, 2$.
 Let $M_1=\mbox{\rm Leaf-Insert}\left(N_1, e_1, n+1\right)$ and 
$M_2 =\mbox{\rm Leaf-Insert}\left(N_2, e_2, n+1\right)$ such that $M_1=M_2$. Then,   there exists a node mapping 
$\phi$ from $M_1$ to $M_2$ such that (i) it maps a leaf in $M_1$ to the same leaf and (ii)
$(\phi(u), \phi(v))\in {\cal E}(M_2)$ if and only if $(u, v)\in {\cal E}(M_1)$. 
Since $n+1$ is inserted as a leaf, $\phi$ maps the parent $p_1$ of $(n+1)$ in $M_1$ to the parent 
$p_2$ of $n+1$ in $M_2$, implying that $\phi$ induces an isomorphic mapping from $N_1$ to $N_2$. This proves the necessity condition. The sufficient condition is straightforward. 

(ii) and (iii)  Both statement can be proved similarly. The proposition is proved.

Figure~\ref{Example_3}
show how to generate the TCN given in Figure~1. Taken together, Propositions ~\ref{prop1}--\ref{prop4} imply the following theorem.

\begin{theorem} 
\label{enumeration_thm}
  Each TCN of ${\cal TCN}_{k+1}(n+1)$ can be obtained from either (i) a unique TCN of ${\cal TCN}_{k+1}(n)$ by attaching Leaf $n+1$ to a tree edge or (ii) a unique TCN  $N\in {\cal TCN}_{k}(n)$ by applying one of the following operations:
  \begin{itemize} 
    \item[{\rm (a)}]   Insertion of  a reticulate node $r$ with the child Leaf $(n+1)$ into a tree edge or straddling two tree edges; 
    \item[{\rm (b)}]  Insert $r$ into a tree edge $(u, v)$,  as described in (a),  and then conduct the child rotation to switch the child of $r$ and the tree node child of $v$. 
   \item[{\rm (c)}]  Insert $r$  straddling two tree edges $e'=(u', v')$ and $e''=(u'', v'')$,  as described in (a),  and  then conduct  the child rotation to switch the child of $r$ and the tree node child of $v''$ (resp. $v'$) if $u''$ (resp. $u'$) is not an ancestor of $u'$ (resp. $u''$). 
  \end{itemize} 
\end{theorem}
%{\bf Proof}.  The statement can be easily derived from Propositions~\ref{prop1}--\ref{prop4} for TCNs.
% The statement for normal networks is based on the fact that normal networks can only be obtained from %normal networks  when TCNs are enumerated via the operations (a)--(c).
%$\QED$

\begin{figure}[b!]
%\begin{figure*}[b!]
            \centering
             \includegraphics[scale = 0.8]{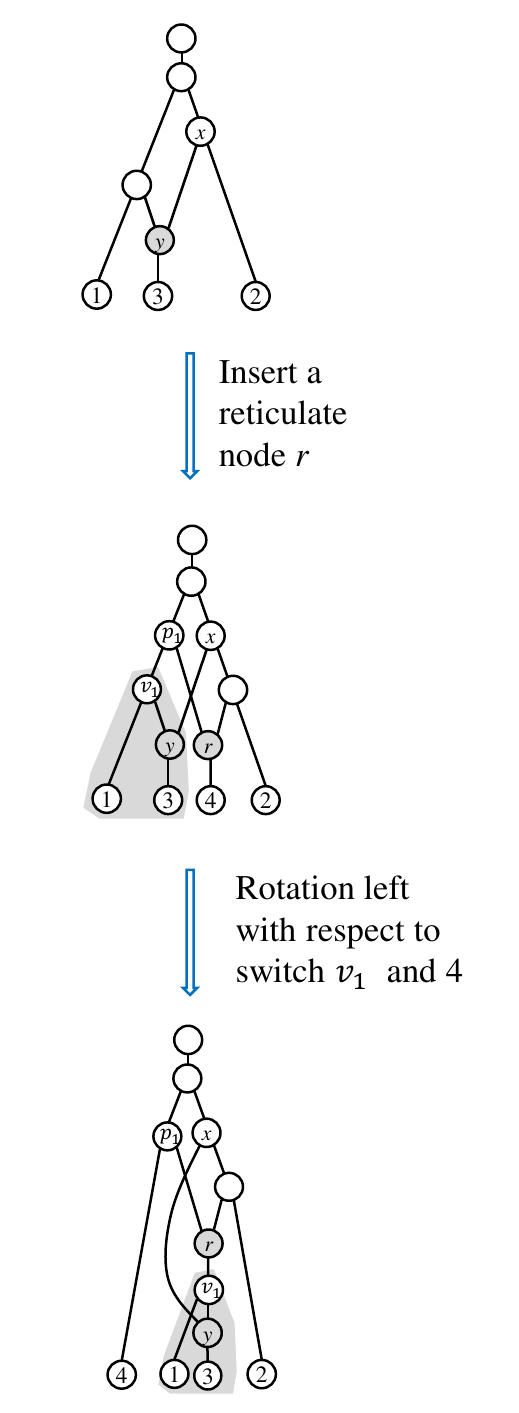}
           \caption{Illustration the undesired condition in Theorem~\ref{enumeration_thm_normal} that prevents from applying s a child rotation. Here, the reticulate node $r$ and its child Leaf 4 are first inserted into the tree edges entering $v_1$ and $v_2$(i.e. 2) in a normal network (left), generating a normal network (middle). But, child rotation to left leads to a tree-child network that is no longer normal (right), in which $y$ does not satisfy the normal condition. }
            \label{Bad_condition}
%\end{figure*}
\end{figure}

If we restrict the operations on normal networks, we obtain all the normal networks in ${\cal TCN}_{k+1}(n+1)$. However, inserting a reticulate node and then applying the child rotation
may lead to a result that  a
reticulation no longer satisfy the normal condition (Figure~\ref{Bad_condition}). Hence, the child-rotation operation should be taken after some verification when all normal networks are enumerated.

\begin{theorem} 
\label{enumeration_thm_normal}
  Each normal network of ${\cal TCN}_{k+1}(n+1)$ can be obtained from either (i)  a unique normal network in ${\cal TCN}_{k+1}(n)$ by attaching Leaf $n+1$ to a tree edge or (ii)   a unique normal network $N\in {\cal TCN}_{k}(n)$ by applying one of the following operations for each pair of
incomparable edges $e_1=(u_1, v_1)$ and $e_2=(u_2, v_2)$ in $N$:
  \begin{itemize} 
    \item[{\rm (a)}]   Insert a reticulate node $r$ with the child $(n+1)$ straddling $e_1$ and $e_2$.
 Let $p_i$ be the tree node inserted into $e_i$ for $i=1, 2$.
    \item[{\rm (b)}]  Insert $r$ as described in (a) and then conduct the child rotation to make  $v_1$ to be the child of $r$ and  $n+1$ the child of $p_1$, respectively, unless 
a reticulate edge $(x, y)$ exists in $N$ (Figure~\ref{Bad_condition}) such that:
  \begin{itemize}
    \item[{\rm (b.1)}]  $y$ is below $v_1$;
    \item[{\rm (b.2)}]  $x$ is not  an ancestor of $v_1$;
   \item[{\rm (b.3)}]  $x$ is an ancestor of $v_2$.
  \end{itemize}
 \item[{\rm (c)}]   Insert $r$ as described in (a) and then conduct the child rotation to make  $v_2$ to be the child of $r$ and  $n+1$ the child of $p_2$, respectively, unless 
a reticulate edge $(x, y)$ exists in $N$ such that:
  \begin{itemize}
    \item[{\rm (c.1)}]  $y$ is below $v_2$;
    \item[{\rm (c.2)}]  $x$ is not  an ancestor of $v_2$;
   \item[{\rm (c.3)}]  $x$ is an ancestor of $v_1$.
  \end{itemize}
  \end{itemize} 
\end{theorem}
{\bf Proof}.  
 The statement for normal networks is based on the fact that if $N$ is obtained from $N'$ vis one of the three operations given in Theorem 1, that the normality of $N$ implies the normality of $N'$. 

The conditions in (b) and (c) are used to exclude the child rotations that make the normal condition invalid for some existing reticulate nodes in the generated TCN.
%$\QED$

\section*{Counting TCNs and normal networks}
\label{sec4}

Let $N$ be a TCN. For a pair of edges $(u_1, v_1)$ and $(u_2, v_2)$ of $N$, they are    {\it incomparable}  if neither of $v_1$ and $v_2$  is an ancestor of the other. 
Let $u(N)$ be the number of unordered pairs of incomparable edges in $N$ and
let:
 \begin{eqnarray}
\label{u_def}
u_{n-1, k-1}=\sum_{N\in {\cal TCN}_{k-1}(n-1)} u(N).
\end{eqnarray}
%We also let $U_{n-1, k-1}$ be the total number of unordered pairs of incomparable edges in all %normal networks in ${\cal TCN}_{k-1}(n-1)$.
%Theorem~\ref{enumeration_thm} implies the following counting result.

Define  $a_{n, k}$ to be $|{\cal TCN}_{k}(n)|$, $0\leq k<n$ and  $b_{n, k}$ to be the number of normal networks in $TCN_{k}(n)$, $0\leq k< n$. 

\begin{theorem} 
\label{cor_1}
(i) 
%Let $a_{n, k}=|{\cal TCN}_{k}(n)|$, $0\leq k<n$. 
% and $b_{n, k}$ denote the number of normal networks in $TCN_{k}(n)$, $0\leq k< n$. 
The $a_{n, k}$ can be calculated through the following recurrence formula:
\begin{eqnarray}
 a_{n,k}&=&
  (2n+k-3)\{a_{n-1,k}+(2n+k-4)a_{n-1, k-1}\} \nonumber\\
  &&+u_{n-1, k-1},
   \label{count_tcn_eqn}
\end{eqnarray}
where  $a_{2, 0}=1$ and $u_{n-1, k-1}$ is defined in Eqn.~(\ref{u_def}).

(ii) The $b_{n, 1}$ can be calculated through the following recurrence formula:
%$$b_{n, n-1}=0,\;\; b_{2, 0}=1,$$
\begin{eqnarray}
\label{count_normal}
b_{n, n-1}=0, \nonumber\\
  b_{n, 1}=%\left\{\begin{array}{ll}
           %0 & \mbox{ if } k=n-1;\\
           % 1 & \mbox{ if } k=0 \mbox{ and } n=2;\\
             (2n-2)b_{n-1, 1}+3u_{n-1, 0} &  n>2,
    %\end{array} \right.
\end{eqnarray}
where $u_{n-1, 0}$ is the total number of unordered pairs of incomparable edges in all the phylogenetic trees on $n-1$ taxa.
\end{theorem}
{\bf Proof.} (i)  The unique tree on two taxa is a TCN and thus $a_{2, 0}=1$.

Each TCN of ${\cal TCN}_{k}(n-1)$ has $2n+k-3$ tree edges and Leaf $n$ can be attached to each of these edges. 
The first term of the right hand side of Eqn.~(\ref{count_tcn_eqn}) counts the TCNs obtained by applying the leaf insertion in Theorem~\ref{enumeration_thm}.

Consider $N\in {\cal TCN}_{k-1}(n-1)$. $N$ has $n-1$ leaves,  $n+k-3$ tree nodes,  and thus 
$2n+k-4$ tree edges.  The reticulation insertion  can be used on a single edge or a pair of edges in $N$. Thus, we can insert a reticulate node $r$ with the child Leaf $n$ in 
$2n+k-4 +{2n+k-4 \choose 2}=(2n+k-3)(2n+k-4)/2$ possible ways. After the insertion of $r$ in a tree edge $(u, v)$, we can apply a child rotation to exchange Leaf $n$ with $v$, as $u$ is not an ancestor of $v$ after $r$ was inserted.  
Similarly, after $r$ is connected  to a pair of edges $e_1$ and $e_2$, we can apply  a child rotation once if one edge is below the other and in two possible ways if neither is an ancestor of the other. 

In summary, for each unordered pair of tree edges $(e_1, e_2)$, we can generate three different tree child networks with $k$ reticulations on $[1, n]$ if they are incomparable and two otherwise. Thus,  we have  the second and third terms of the formula.  

(ii) The fact that $b_{n, n-1}=0$ was first proved by Bickner \cite{Bicker_12}.   

%It is trivial that $a_{2, 0}=1$.

In the case that $n>2$ and $k\leq n-2$, Eqn.~(\ref{count_normal}) for $b_{n, 1}$ follows from the following two facts: 
\begin{itemize}
  \item Only two incomparable edges in normal networks in
 ${\cal TCN}_{k-1}(n-1)$ can be used to generate normal networks in ${\cal TCN}_{k}(n)$; 
\item For each unordered pair of incomparable edges in a tree on $[1, n-1]$, three normal networks can be obtained by applying insertion of reticulate node and two child rotations. 
\end{itemize}
% $\QED$

Unfortunately, we do not know how to obtain a simple formula for $b_{n, k}$ in general. 
By Theorem~\ref{cor_1}, one still can compute the number of normal networks with $k$ reticulate nodes on $[1, n]$, $b_{n,k}$, by enumeration. 
For each  $1\leq k \leq n-2$ and $3\leq n\leq 7$, $b_{n, k}$ is listed in Table~\ref{Table1}. 

\setlength{\tabcolsep}{5pt}
{\small
\begin{table}
%\centering 
\caption{ [Table 1 is here]
%Counts of  the normal networks with $k$ reticulations on $[n]$, $1\leq k \leq n-2$ 
%and $3\leq n\leq 7$. 
\label{Table1}
}
\begin{tabular}{c|rrrrrr}
 \hline
  $k$\textbackslash $n$ &    3 & 4 & 5 & 6 & 7 & 8 \\ 
\hline
  1   & 3  & 54 & 855       & 14,040 & 248,535 & 4,787,370 \\
  2   &    &  48 & 2,310    & 78,120 & 2,377,620 & 70,749,000 \\
  3 &      &      & 1,920 &  184,680 & 11,038,530 & 536,524,830\\
  4 &      &     &         &  146,520 & 23,797,302 & 2,217,404,379\\
5  &      &      &         &              & 16,198,764 &3,802,965,091 \\
6  &     &   &   &       &                                 &2,479,006101\\
\hline
\end{tabular}
\end{table}
}

 It is challenging to obtain a simple formula for counting $u_{n, k}$ for arbitrary $k$.  But we can find a closed formula for $u_{n, 0}$ and thus obtain a recurrence formula for  $a_{n, 1}$ and $b_{n, 1}$.

\begin{lemma}
\label{no_incomparable_pairs}
 For any $n\geq 2$,  the total number of unordered pairs of incomparable edges in all the phylogenetic trees on $n$ taxa is: 
 \begin{eqnarray}
  u_{n, 0}=\frac{(n+1)(2n)!}{2^{n}(n)!} -2^{n}n! \label{incomparable_edge}
\end{eqnarray}
\end{lemma}
{\bf Proof.} 
Let $T$ be a phylogenetic tree on $[1, n-1]$ and let 
$O(T)$ denote the set of ordered pairs of comparable edges in $T$, where that $(x, y)\in O(T)$ means 
the edge $x$ is above the edge $y$. Then, it is not hard to verify: 
\begin{eqnarray*}
 O(T)&=&\cup_{e\in {\cal E}(T)} 
   \{(e, x) \;|\;  x\in {\cal E}(T) \mbox{ s.t. $e$ is above $x$}\}\\
   &=&\cup_{e\in {\cal E}(T)} 
   \{(y, e) \;|\;  y\in {\cal E}(T) \mbox{ s.t. $e$ is below $y$}\}
\end{eqnarray*}

 Assume $T'$ is obtained from $T$ by attaching Leaf $n$ in an edge $e=(u, v)$. In $T'$, the parent $w$ of Leaf $n$ is the tree node inserted in $e$, implying that  $e$ is subdivided into two edges of $T'$:
  $$e_1=(u, w), \;\;e_2=(w, v),$$
and 
$${\cal E}(T')=\{e_1, e_2, (w, n)\}\cup {\cal E}(T) - \{e\}. $$
Thus, 
{\small 
\begin{eqnarray*}
 O(T') &=&  \{(e_1, e_2), (e_1, (w, n))\} \\
    && \cup \{(x, y) \in O(T) \;|\; x\neq y, x\neq e \neq y \}\\
    & & \cup \{ (e', e_1), (e', e_2), (e', (w, n)) \;|\; (e', e)\in O(T)\} \\
    & & \cup \{( e_1, e''), (e_2, e'') \;|\; (e, e'')\in O(T)\}.
\end{eqnarray*}
}
Hence, 
{\small
\begin{eqnarray*}
&& |O(T')|\\
&= &|O(T)|+2+2|\{(x, e) \;|\;  x\in {\cal E}(T): (x, e)\in O(T) \}|\\
          && +     |\{(e, y) \;|\;  y\in {\cal E}(T) \mbox{ s.t. $y$ is below $e$}\}|.
\end{eqnarray*}
}
Since $T$ has $2n-3$ edges, 
\begin{eqnarray}
  \sum_{T'\in \mathcal{LI}(T, n)}|O(T')| &=&(|{\cal E}(T)|+3) |O(T)|+2 |{\cal E}(T)| \nonumber \\
   &=&2n\times O(T)+2(2n-3). \label{last}
\end{eqnarray}
where $\mathcal{LI}(T, n)$ denotes the set of  $2n-3$ phylogenetic trees  that are obtained by a Leaf-Insertion on $T$. 

Let $c_n$ be the total number of 
unordered pairs of comparable edges in all the phylogenetic trees on $n$ taxa. Clearly, 
$c_2=2$.  Since there are 
$\frac{(2n-4)!}{2^{n-2}(n-2)!}$ phylogenetic trees with $n-1$ leaves,  which each have $2n-3$ edges, 
Eqn. (\ref{last}) implies:
\begin{eqnarray*}
   c_{n}&=&2nc_{n-1}+\frac{2(2n-2)!}{2^{n-1}(n-1)!}
\end{eqnarray*}
or, equivalently, 
\begin{eqnarray*}
   \frac{1}{n!}c_{n}&=&2\left(\frac{1}{(n-1)!}c_{n-1}\right)+\frac{(2n-2)!}{2^{n-2}n!(n-1)!}.
\end{eqnarray*}
Therefore, 
\begin{eqnarray*}
  %  & =& 2n\cdot 2(n-1) \cdot 2(n-k) c_{n-k-1} + \sum^k_{i=0}\frac{2^{n-k-i}n!}{(n-k-i)!}|\\
%  & = & 2^{n-2} n! + \sum^{n-3}_{k=0} \frac{2^{k+1}n!}{(n-k)!}\frac{(2n-2k-2)!}{2^{n-k-1}(n-k-1)!}\\
% c_{n} &=&   \sum^{n-2}_{k=0} \frac{2^{k+1}n!}{(n-k)!}\frac{(2n-2k-2)!}{2^{n-k-1}(n-k-1)!}\\
%  &=& \sum^{n-1}_{k=1} \frac{2^{k}n!}{(n-k+1)!}\frac{(2n-2k)!}{2^{n-k}(n-k)!}\\
% &=&{n!2^n} \sum^{n-1}_{k=1} \frac{1}{(n-k+1)!}\frac{(2n-2k)!}{2^{2(n-k)}(n-k)!}\\
 c_n &=& {n!2^n} \sum^{n-1}_{k=1} \frac{1}{(k+1)!}\frac{(2k)!}{2^{2k}(k)!}\\
 &=& \frac{n!2^{n-1}}{\pi} \sum^{n-1}_{k=1} \int^{4}_{0}\left(\frac{x}{4}\right)^k\left(\frac{4-x}{x}\right)^{1/2}dx\\
% &=&\frac{n!2^{n-1}}{\pi} \sum^{n-1}_{k=1} \int^{4}_{0}\left(\frac{x}{4}\right)^k\left(\frac{4-x}{x}\right)^{1/2}dx\\
%&=&\frac{n!2^{n+1}}{\pi} \sum^{n-1}_{k=1} \int^{1}_{0}x^k\left(\frac{1-x}{x}\right)^{1/2}dx\\
%&=&\frac{n!2^{n+1}}{\pi}  \int^{1}_{0}\frac{x(1-x^{n-1})}{1-x}\left(\frac{1-x}{x}\right)^{1/2}dx\\
&=& \frac{n!2^{n+1}}{\pi} \int^{1}_{0} (1-x^{n-1})\left(\frac{x}{1-x}\right)^{1/2}dx\\
%&=& 2^n n!\left(1- \frac{(2n)!}{2^{2n-1}(n!)(n!)}\right)\\
&=& 2^n n! - \frac{(2n)!}{2^{n-1}n!},
\end{eqnarray*}
where $\frac{(2k)!}{(k+1)!k!}$ is the $k$-th Catalan number $C_k$ that is equal to the integral appearing above (\cite{Richard_SBook}).
Since there are $\frac{(2n-2)!}{2^{n-1}(n-1)!}$ phylogenetic trees on $[1, n]$ each  having $(2n-1)$ edges, 
{\small 
\begin{eqnarray*}
 u_{n, 0}&=&\frac{(2n-2)!}{2^{n-1}(n-1)!} {2n-1 \choose 2} -c_n\\
          & = & \frac{(2n-2)!}{2^{n-1}(n-1)!} (n-1)(2n-1) + \frac{(2n)!}{2^{n-1}n!} - 2^n n! \\
         & = & \frac{(n+1)(2n)!}{2^{n}n!} -2^n n!.
\end{eqnarray*}
}
%$\QED$ 

\begin{theorem}
  For any $n\geq 3$, the numbers of TCNs and normal networks with exactly one reticulate node
on $n$ taxa are:
 \begin{eqnarray}
\label{x_count}
   a_{n, 1}
&= &  \frac{(2n)!}{2^n (n-1)!}-2^{n-1} n!
\end{eqnarray}
and
\begin{eqnarray}
\label{n_count}
  b_{n, 1} =\frac{(n+2)(2n)!}{2^{n}n!}-3\cdot 2^{n-1}n!,
\end{eqnarray}
respectively.
\end{theorem}
{\bf Proof.} Since $a_{n-1, 0}=\frac{(2n-4)!}{2^{n-2}(n-2)!}$,  by Theorem~\ref{cor_1}, 
{\small
\begin{eqnarray*}
 a_{n, 1}=2(n-1)a_{n-1, 1}+\frac{(3n-2)(2n-3)!}{2^{n-2}(n-2)!}-2^{n-1}(n-1)!
\end{eqnarray*}
}
or, equivalently,
{\small 
\begin{eqnarray*}
 \frac{a_{n, 1}}{(n-1)!}=2\left(\frac{a_{n-1, 1}}{(n-2)!}\right)+\frac{(3n-2)(2n-2)!}{2^{n-1}((n-1)!)^2}-2^{n-1}
\end{eqnarray*}
}
Therefore, since $a_{2, 1}=2$, 
\begin{eqnarray*}
&& \frac{a_{n, 1}}{(n-1)!}\\
&=& 2^{n-2} \left(\frac{a_{2, 1}}{(2-1)!}\right)\\
   &&+ \sum^{n-2}_{i=1} \frac{(3n-3i+1)(2n-2i)!}{2^{n-2i+1}((n-i)!)^2} - 2^{n-1} (n-2) \\
  &=& \sum^{n-2}_{i=1} \frac{(3n-3i+1)(2n-2i)!}{2^{n-2i+1}((n-i)!)^2} - 2^{n-1} (n-3) \\
   &=& 2^{n-1}\sum^{n-2}_{i=1} \frac{(3(n-i)+1)(2(n-i))!}{2^{2(n-i)}((n-i)!)^2} - 2^{n-1} (n-3) \\
  &=& 2^{n-1}\sum^{n-1}_{k=2} \frac{(3k+1)(2k)!}{2^{2k}(k!)^2} - 2^{n-1} (n-3) \\
 &=& 2^{n-1}\sum^{n-1}_{k=1} {2k\choose k} \frac{3k+1}{4^{k}} - 2^{n-1} (n-1).
\end{eqnarray*}
By induction, we can show that $\sum^{n}_{k=0}{2k\choose k}4^{-k}=\frac{(2n+1)!}{2^{2n}(n!)^2}$ and
$\sum^{n}_{k=0}{2k\choose k}k4^{-k}=\frac{(2n+1)!}{3\cdot 2^{2n}n!(n-1)!}$. Continuing the above calculation, we obtain:
{\small
$$\frac{a_{n, 1}}{(n-1)!}=\left\{\frac{(2n-1)!}{2^{(n-1)}(n-1)!}\left[\frac{1}{(n-2)!}+\frac{1}{(n-1)!}\right]-n\right\}.$$
}

This proves Eqn.~(\ref{x_count}). 

%\newpage
Similarly, by Theorem~\ref{cor_1} and Lemma~\ref{no_incomparable_pairs},  we have:
\begin{eqnarray*}
 b_{n, 1}&=& 2(n-1)b_{n-1, 1} \\
   & &+ 3\cdot \left(\frac{n(2n-3)!}{2^{n-2}(n-2)!}-2^{n-1}(n-1)!\right)\\
\end{eqnarray*}
or, equivalently,
\begin{eqnarray*}
 \frac{b_{n, 1}}{(n-1)!}&=& 2\frac{b_{n-1, 1}}{(n-2)!} + 3 {2n-3\choose n-1}\frac{n}{2^{n-2}}-3\cdot 2^{n-1}.
\end{eqnarray*}
Since $b_{2, 1}=0$, 
%{\small
\begin{eqnarray*}
 &&\frac{b_{n, 1}}{(n-1)!}\\
   &=& 2^{n-2}\frac{b_{2, 1}}{(2-1)!}\\
   && +3\cdot 2^n \cdot \sum^{n-2}_{i=1} {2n-2i-1\choose n-i}\frac{n-i+1}{2^{2n-2i}}\\
   && -3\cdot 2^{n-1} (n-2)\\
   &=& 3\cdot 2^n \cdot \sum^{n-1}_{k=1} {2k-1\choose k}\frac{k+1}{2^{2k}}-3\cdot 2^{n-1} (n-1)\\
 &=& 3\cdot 2^{n-1} \cdot \sum^{n-1}_{k=1} {2k \choose k}\frac{k+1}{2^{2k}}-3\cdot 2^{n-1} (n-1)\\
&=&\frac{(n+2)(2n)!}{2^{n}n!}-3\cdot 2^{n-1}n!.
\end{eqnarray*}
%}
This proves Eqn.~(\ref{n_count}). %$\QED$

\noindent {\bf Remark}   Every RPN with exactly one reticulate node  is  a TCN. Therefore,   $a_{n, 1}$ is actually the number of RPNs with one reticulate node. 

\section*{Conclusions}

It is well-known that all phylogenetic trees on $n$ taxa can be generated by the insertion of the $n$-th taxa in each edge of
all the phylogenetic trees on the first $n-1$ taxa.  The main result of this work is a generalization of this fact into TCNs.  This leads to a simple procedure for enumerating  both normal networks and TCNs, the C-code for which  is available upon request.  It is fast enough to count all the normal networks on eight taxa.  Recently,  Cardona {\it et al.} introduced a novel operation to enumerate TCNs. Their program was successfully used to compute the exact number of tree-child networks on six taxa. Although our program is faster than theirs, it still cannot be used to count TCNs on eight taxa on a PC.   
%Whether or not there exists a closed formula for counting TCNs on $n$ taxa remains open.  

Another contribution of this work is  Eqn.~(\ref{x_count})  and (\ref{n_count}) for counting RPNs with exactly one reticulate node.  Semple and Steel \cite{Steel_06}  presented formulas for counting unrooted networks with one reticulate node. Since an unrooted networks can be oriented into a different number of RPNs,  it is note clear how to use their results to derive a formula for the count of RPNs.  Bouvel et al. \cite{Bouvel_18} presented a formula for counting RPNs with one reticulate node. Our formula is much simpler than the formula given in \cite{Bouvel_18}. 

Lastly, the following problem is open:

Is there a simple formula like Eqn.~(\ref{x_count}) for the count of TCNs with $k$ reticulate nodes on $n$ taxa for each $k>1$?

%\section*{Section title}
%Text for this section \ldots
%\subsection*{Sub-heading for section}
%Text for this sub-heading \ldots
%\subsubsection*{Sub-sub heading for section}
%Text for this sub-sub-heading \ldots
%\paragraph*{Sub-sub-sub heading for section}
%Text for this sub-sub-sub-heading \ldots

%%%%%%%%%%%%%%%%%%%%%%%%%%%%%%%%%%%%%%%%%%%%%%
%%                                          %%
%% Backmatter begins here                   %%
%%                                          %%
%%%%%%%%%%%%%%%%%%%%%%%%%%%%%%%%%%%%%%%%%%%%%%

\begin{backmatter}

\section*{Competing interests}
  The author declares that he has no competing interests.

\section*{Author's contributions}
    The author conducted research and wrote the manuscript.

\section*{Acknowledgements}
  The author thanks Mike Steel for useful discussion of counting networks and comments on  the manuscript and Yurui Chen for implementing the implmentation of the enumeration method given in this work.
He also thanks anonymous reviewers and Jonathan Klawitter for useful feedback on the first draft of this paper.

\section*{Funding}
This work was partially supported by Singapore's Ministry of Education Academic
Research Fund Tier-1 [grant R-146-000-238-114] and the National Research
Fund [grant NRF2016NRF-NSFC001-026].

%%  After submission of the .TEX file,                     %%
%%  you will be prompted to submit your .BBL file.         %%

% if your bibliography is in bibtex format, use those commands:
\bibliographystyle{bmc-mathphys} % Style BST file (bmc-mathphys, vancouver, spbasic).
\bibliography{Normal_networks_article}      % Bibliography file (usually '*.bib' )
% for author-year bibliography (bmc-mathphys or spbasic)
% a) write to bib file (bmc-mathphys only)
% @settings{label, options="nameyear"}

% or include bibliography directly:
% \begin{thebibliography}
% \bibitem{b1}
% \end{thebibliography}

%%%%%%%%%%%%%%%%%%%%%%%%%%%%%%%%%%%
%%                               %%
%% Figures                       %%
%%                               %%
%% NB: this is for captions and  %%
%% Titles. All graphics must be  %%
%% submitted separately and NOT  %%
%% included in the Tex document  %%
%%                               %%
%%%%%%%%%%%%%%%%%%%%%%%%%%%%%%%%%%%

%%
%% Do not use \listoffigures as most will included as separate files

\end{backmatter}
\end{document}